\documentclass[journal=jacsat,manuscript=article]{achemso}
\usepackage{color}
\usepackage{amsmath,amssymb}
\usepackage{graphicx}
\usepackage{float}
\usepackage{natbib}
\usepackage{color}
\usepackage[colorlinks=true,bookmarks=false,citecolor=blue,urlcolor=blue]{hyperref}
\usepackage{lmodern}
\usepackage[most]{tcolorbox}
\usepackage{tikz}
\usepackage{varwidth}
\usepackage{wrapfig}
\usepackage{capt-of}
\usepackage{paracol}
\usepackage{soul}
\usepackage{xcolor}
\usepackage[normalem]{ulem}
\usepackage[font=doublespacing, justification=raggedright]{caption}

\newcommand{\rev}{\textcolor{black}}

\usepackage[version=3]{mhchem} 

\author{Abdulmalik A. Madigawa}
\altaffiliation{Contributed equally to this work}
\affiliation{\dtu}

\author{Jan N. Donges}
\altaffiliation{Contributed equally to this work}
\affiliation{\tub}

\author{Benedek Gaál}
\affiliation{\dtu}

\author{Shulun Li }
\affiliation{\tub}
\altaffiliation{\cas}
\altaffiliation{\ucas}

\author{\rev{Martin Arentoft Jacobsen}}
\affiliation{\dtu}

\author{Hanqing Liu }
\altaffiliation{\cas}
\affiliation{\ucas}

\author{Deyan Dai}
\altaffiliation{\cas}
\affiliation{\ucas}

\author{Xiangbin Su}
\altaffiliation{\cas}
\affiliation{\ucas}

\author{Xiangjun Shang}
\altaffiliation{\cas}
\affiliation{\ucas}

\author{Haiqiao Ni}
\altaffiliation{\cas}
\affiliation{\ucas}

\author{Johannes Schall}
\affiliation{\tub}

\author{Sven Rodt}
\affiliation{\tub}

\author{Zhichuan Niu}
\altaffiliation{\cas}
\affiliation{\ucas}

\author{Niels Gregersen}
\affiliation{\dtu}

\author{Stephan Reitzenstein}
\email{stephan.reitzenstein@physik.tu-berlin.de}
\affiliation{\tub}

\author{Battulga Munkhbat}
\email{bamunk@dtu.dk}
\affiliation{\dtu}

\newcommand{\dtu}{
    Department of Electrical and Photonics Engineering, Technical University of Denmark,
    2800 Kgs. Lyngby, Denmark
}
\newcommand{\tub}{
    Institute für Festkörperphysik,
    Technische Universität Berlin,
    10623 Berlin, Germany
}

\newcommand{\cas}{
State Key Laboratory for Superlattice and Microstructures, Institute of Semiconductors, Chinese Academy of Sciences, Beijing 100083, China 
}

\newcommand{\ucas}{
Center of Materials Science and Optoelectronics Engineering, University of Chinese Academy of Sciences, Beijing 100049, China  
}

\title
  {Assessing the alignment accuracy of state-of-the-art deterministic fabrication methods for single quantum dot devices}

 \keywords{quantum dots, localization, photoluminescence, cathodoluminescence, QD imaging, deterministic fabrication, single-photon source.}

\begin{document}

\newpage
\begin{abstract}

The realization of efficient quantum light sources relies on the integration of self-assembled quantum dots (QDs) into photonic nanostructures with high spatial positioning accuracy.
In this work, we present a comprehensive investigation of the QD position accuracy, obtained using two marker-based QD positioning techniques, photoluminescence (PL) and cathodoluminescence (CL) imaging, as well as using a marker-free in-situ electron beam lithography (in-situ EBL) technique. We employ four PL imaging configurations with three different image processing approaches and compare them with CL imaging. We fabricate circular mesa structures based on the obtained QD coordinates from both PL and CL image processing to evaluate the final positioning accuracy. This yields final position offset of the QD relative to the mesa center of $\mu_x$ = (-40$\pm$58) nm and $\mu_y$ = (-39$\pm$85) nm with PL imaging and $\mu_x$ = (-39$\pm$30) nm and $\mu_y$ = (25$\pm$77) nm with CL imaging, which are comparable to the offset $\mu_x$ = (20$\pm$40) nm and $\mu_y$ = (-14$\pm$39) nm obtained using the in-situ EBL method. We discuss the possible causes of the observed offsets, which are significantly larger than the QD localization uncertainty obtained from simply imaging the QD light emission from an unstructured wafer. Our study highlights the influences of the image processing technique and the subsequent fabrication process on the final positioning accuracy for a QD placed inside a photonic nanostructure.

\end{abstract}

\maketitle

\section{Introduction}

Solid-state single-photon emitters are crucial building blocks for developing efficient quantum light sources and on-chip quantum circuits for quantum information processing platforms. Self-assembled semiconductor quantum dots (QDs) are one of the most promising solid-state quantum emitters for realizing quantum communication networks \cite{Gisin2007QuantumCommunication,Vajner2021QuantumDots,Wehner2018QuantumAhead}, photonic quantum computation\cite{Kok2007LinearQubits, Madsen2022QuantumProcessor, Wang2019BosonSpace} which will enable many applications in photonic quantum technologies \cite{Heindel2023QuantumTechnology}. Here, one uses the fact that a single QD efficiently emits single photons due to its quantized two-level electronic structure. However, extracting and collecting the emitted photons for usable application is rather challenging, and in a simple planar device geometry, most of the photons are restrained in the semiconductor matrix due to total internal reflection. This problem is usually tackled by incorporating the QD within an engineered photonic nanostructure \cite{Claudon2010ANanowire,Arcari2014Near-UnityWaveguide,Wang2020a,Liu2019AIndistinguishability,Ding2016,Schnauber2019IndistinguishableCircuits,Gschrey2015HighlyLithography,Wei2014DeterministicPassage}. In cavity-based quantum light sources, this device configuration allows for efficient coupling into only one single photonic mode by the Purcell effect, maximizing the photon extraction efficiency. Importantly, maximum Purcell enhancement is attained when the QD is spatially positioned at the mode's electric field maximum and in spectral resonance with the cavity mode.  While there has been success in designing high-performance nanophotonic components for efficient light extraction, one very challenging aspect is the accurate spatial and spectral positioning of a single QD within the structures. In fact, the underlying self-assembled growth process results in random positions of the QDs. In addition, the QDs have different shapes and material compositions, leading to variations in their emission wavelengths. The described randomness poses a significant challenge in fabricating quantum light sources with spatially and spectrally resonant QDs, which is crucial for optimum device performance 

\rev{The spatial displacement of the QD from the center of nanostructures has led to a large discrepancy between the theoretically expected and experimentally measured Purcell factors, especially for devices with a QD in close proximity to the surface of the structure \cite{Sapienza2015,Nawrath2023BrightC-Band,Rickert2019OptimizedGratings}. } 
\rev{Furthermore, recently, it has been found that in the case of circular Bragg grating (CBG) cavities, a displacement between QD and photonic structure of 100 nm can lead to a maximum polarization anisotropy, which limits their usage for generation of flying qubits based on the polarization degree of freedom such as polarization-entangled photon pairs \cite{Peniakov2023PolarizedResonator}. } 
\rev{In addition, to fabricate optimal devices} for scaling up to larger quantum photonic systems such as large-scale integrated quantum photonic circuits \rev{\cite{Sartison2022ScalableCircuits}, it is necessary to reliably integrate quantum emitters into nanobeam or photonic crystal cavities with an alignment accuracy better than 50 nm to maintain appropriately high photon coupling efficiency and light-matter interaction.}

Over the years, different technology platforms have been developed to select and determine the position and spectrum of target QDs and to integrate them deterministically into nanophotonic structures acting as high-performance quantum light sources. Methods like in-situ photolithography\cite{Dousse2008,Sartison2017CombiningSpectroscopy,Kolatschek2019DeterministicLithography}, in-situ e-beam lithography\cite{Gschrey2013}, and photoluminescence imaging \cite{Thon2009StrongCavity,Kojima2013AccurateImaging,Sapienza2015,He2017,Liu2017c,Pregnolato2020DeterministicDots,Liu2021a} have been developed and employed very successfully for deterministic QD-device processing. The field was pioneered by the development of in-situ optical lithography in 2008, which involves the QD localization (determining the QD location) with photoluminescence (PL) mapping and subsequent optical lithography of the photonic structure. This method can achieve localization accuracy within $\pm$50 nm and has been employed in developing highly efficient micropillar single-photon sources\cite{Somaschi2016}. However, in-situ optical lithography can only be applied to specific geometries like micropillars. Moreover, the fabrication is limited to structures with feature sizes achievable within the optical diffraction limit. On the other hand, the in-situ electron beam lithography (in-situ EBL) approach that combines low-temperature cathodoluminescence (CL) mapping for spatial and spectral QD selection with high-resolution electron beam lithography (EBL) has proven to be very powerful in terms of lithography resolution and geometrical flexibility. The pixel size and resolution can be picked freely and is not fixed by a camera sensor. Furthermore, the high electron energy enables efficient excitation independent of the semiconductor bandgap and does not overlay with the QD luminescence signal. The in-situ EBL is known to feature high alignment accuracy of 30-40 nm and was initially employed for the development of different high-performance single-photon sources \cite{Gschrey2015HighlyLithography}. Quite naturally, this advanced nanotechnology platform can also be used to define complex patterns, such as integrated quantum photonic circuits \cite{Schnauber2018DeterministicLithography,Schnauber2019IndistinguishableCircuits,Li2023ScalableCircuit}. However, it is technically rather complex as it requires the combination of low-temperature CL spectroscopy with low-temperature EBL in a single system. As still a further deterministic nanofabrication technology, the PL imaging approach has become very popular in recent years. It involves the QD localization and the structure fabrication in separate process steps. In this approach, high QD preselection accuracy as low as 5 nm has been reported~\cite{Liu2017}. However, the process flow is more complex than in the in-situ lithography techniques. This is explained by the fact that it is a marker-based localization approach and involves the lithography step in a different setup, which potentially results in bigger errors in aligning the nanophotonic structures to the preselected QDs. Several studies report high localization accuracy from PL image analysis, but so far, no detailed analysis of the final alignment uncertainty has been performed, including the aforementioned in-situ lithography techniques. 

\begin{figure}
\includegraphics[width=0.4 \textwidth]{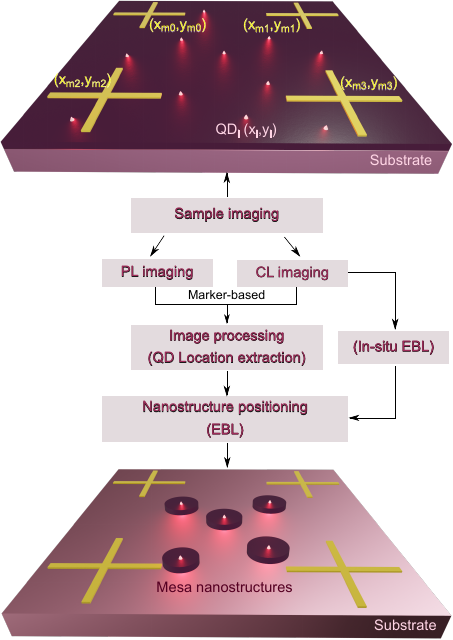}
\caption{Procedure for nanostructures positioning around pre-selected QDs. Images are taken using either PL or CL imaging systems and processed using rigorous image analysis algorithms to precisely extract the location of the target QDs relative to alignment markers. The determined coordinates are then used to fabricate nanostructures (here, circular mesas) around the QDs using EBL. The In-situ EBL combines CL imaging for marker-free QD localization with EBL structuring in one process setup.}
\label{Figure 1}
\end{figure} 

Here, we report on a systematic and comparative study of the final alignment accuracy of a target QD in a photonic structure using marker-based PL and CL imaging techniques. Furthermore, we compare the two marker-based techniques with the marker-free in-situ EBL technique, with the aim of uncovering the distinct advantages and disadvantages associated with each method. Our approach starts with the determination of the QD position using the different QD imaging techniques through a series of image processing steps. We then fabricate a circular mesa test structure with a suitably large diameter around the localized QD. Finally, we extract the final position of the QDs within the mesa structures by mapping the QD emission. We study the mean offsets (deviations from the center of the mesa) and uncertainties in the QD position for the different techniques. All approaches exhibit notable alignment offsets and uncertainties much larger than those obtained from the simple image-processing fit uncertainties, which are often used as a measure of the alignment accuracy of deterministic QD device processing. We analyze the errors in each approach and propose strategies to improve the QD positioning accuracy. 

\section{QD localization}

The QD heterostructure sample we use in this study is grown using molecular beam epitaxy (MBE) and consists of low-density InAs QDs embedded in a GaAs membrane. An array of alignment markers is fabricated on the planar sample using EBL, followed by Ti/Au deposition and a lift-off process. The alignment markers are a set of four square-cross marks with an arm length of 30 $\mu$m and a width of 2 $\mu$m. They serve as reference markers for extracting the global coordinates of the QDs for further EBL processing. To compare the different positioning techniques, QDs from the same sample regions are selected, and the QD coordinates are determined using the described techniques, including marker-based PL and CL and in-situ EBL. The determined coordinates are then used in fabricating mesa structures around the QDs (Figure \ref{Figure 1}), with the QDs divided between the different techniques to evaluate the accuracy of each separately.\\

\noindent \textbf{QD localization via PL imaging}

First, we use the PL imaging technique to determine the center position of selected QDs with reference to the alignment marks. The setup employed for the sample imaging is based on the two-color PL imaging technique developed in Ref.~\citenum{Sapienza2015} (see Supporting Information). The sample is mounted in a closed-cycle cryostat operating at 4 K. The cryostat is equipped with a piezoelectric positioning stage and a low-temperature microscope objective (magnification = 60$\times$, NA = 0.82) located inside the cryostat. The PL setup utilizes two different color light-emitting diodes (LEDs) for the excitation and imaging of the QDs and alignment markers. A 470 nm LED is used for QD excitation, while a 1050 nm LED is used for imaging the alignment markers. The wavelengths of the LEDs are carefully chosen to achieve optimal contrast for both the markers and the QDs while minimizing emission and reflection from the sample background. To extract the QDs location coordinates, the QD PL and alignment marker images are taken simultaneously using a CMOS camera (2048 pixels x 2040 pixels resolution) within an $\approx$ (86 x 86) $\mu$m$^2$ field of view with an image acquisition time of 1 s. The images are processed using an image analysis program developed in Python (see Supporting Information for details). The program utilizes a combination of the cross-correlation algorithm to locate the center of the four markers\cite{anderson2004}, and the Gaussian blob detection with maximum likelihood estimator (MLE) algorithm to locate the QDs\cite{mortensen2010}. The center coordinates of the markers and the QDs are retrieved in pixel units and transformed into local coordinates, with the top-left marker serving as the origin. The local coordinates are then transformed into global EBL coordinates for the subsequent fabrication of nanostructures. Furthermore, after the mapping step, the PL spectrum of each selected QD is recorded and analyzed to confirm the presence of a single QD. The 2D intensity profiles of the QDs are fitted with a Gaussian function using a nonlinear least-squares approach, and the position uncertainties are extracted from the peak position error of the fit, represented as one standard deviation. The position uncertainty of the markers is extracted from the polynomial fits of a line cut along the cross-correlation maximum (with a 68\% confidence interval).

\begin{figure}
\includegraphics[width=1\textwidth]{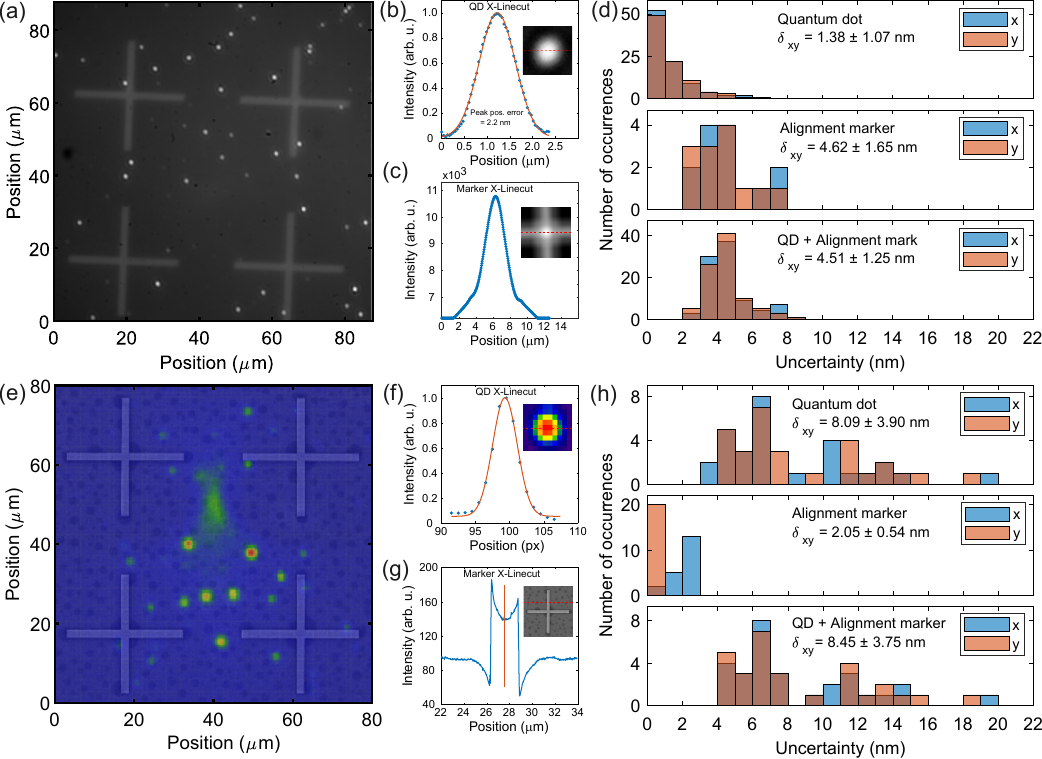}
\caption{Determining the location of QDs with reference to alignment markers using QD imaging techniques. (a)-(d) displays the two-color PL method (Single-image 1) and (e)-(h) the CL method. Image obtained using (a) PL imaging setup and (e) CL mapping setup. (b) Intensity line cut profile (x-axis) of QD PL (blue dots) and its Gaussian fit (red line, with one standard deviation peak position error) and (c) intensity line cut profile (x-axis) of cross-correlated marker image along the located center, of the image in (a). (d) Histogram of the uncertainties in the QDs location, alignment markers location, and the combined uncertainties of the QD and marker (QD+marker) of the image in (a), measured from the line cuts from 15 images (taken at different field regions on the sample). (f) and (g) Intensity line cut (x-axis) of QD CL and marker image, respectively, of the image in (e). (h) Histogram of the uncertainties in the QD location, alignment marker location, and the combined uncertainties of the QD and marker (QD + marker) of the image in (e). The location uncertainties of the QDs were extracted from the 2D Gaussian fit of the QDs profiles for both techniques. The location uncertainties of the alignment markers in the PL technique were extracted from the polynomial fit of a cropped region of interest (20 x 20-pixel area) around the cross-correlation center (with a 68\% confidence interval), while the CL marker uncertainties were determined from a straight line fit through the marker center. Combined QD+ alignment marker uncertainties are obtained by propagating the uncertainties of the QDs and alignment markers.}
\label{figure 2}
\end{figure}

To identify the most effective approach for accurate localization of the QDs in relation to the markers, we acquire and analyze four sets of images obtained using different imaging settings and configurations. Two sets of images are acquired using the two-color PL imaging approach, each with different contrast settings (Single-image 1 and Single-image 2). The image contrast of the QDs and alignment markers plays a critical role in achieving precise localization of the QDs during image processing. Optimal contrast is achieved by carefully adjusting the illumination and excitation LED powers. Figure \ref{figure 2}(a)  shows an image taken with the power level of the 1050 nm and 470 nm LEDs adjusted to make the QD brighter than the alignment markers (Single-image 1). This adjustment effectively reduces the background reflection from the 1050 nm LED, improving the QD contrast. An analysis of the uncertainty for the QD and marker locations, as depicted in Figure \ref{figure 2}(d), reveals a mean uncertainty of (1.38 $\pm$ 1.07) nm for the QD location and (4.62 $\pm$ 1.65) nm for the marker location, leading to a combined uncertainty (QD+marker) of (4.51 $\pm$ 1.25). The QD+ marker uncertainty is obtained by adding the QD and marker uncertainties using the error propagation formula. The low uncertainty for both the markers and the QDs can be attributed to the enhanced signal-to-noise ratio achieved by utilizing a low 1050 nm LED illumination power, which is crucial for minimizing fitting errors. The QD uncertainties depicted here are extracted from a 2D Gaussian fit of the QD intensity profile. The uncertainty extracted from the MLE method is as low as 0.5 nm. However, this value is unreliable as it depends strongly on the QD pixel intensity, which varies from image to image (see Supporting Information note S3)

The effect of the marker and QD contrast in the overall uncertainty of the QD position for the PL imaging approach is studied with different imaging configurations (see Table \ref{Table 1}). Single-image 2 is acquired using the same imaging configuration as Single-image 1, with the power level of the 1050 nm LED adjusted to capture a bright marker image and enhance the marker contrast (see Supporting Information for images and histograms of uncertainties). The improvement in marker contrast results in better accuracy in the marker location as compared to Single-image 1. However, an increase in the uncertainty of the QD location is observed. This is attributed to the low QD's luminescence contrast caused by the increased background reflection of the illumination LED. In addition, two sets of images are also acquired using a single-color approach (one LED at a time). The first set (Merged-images) involves acquiring two separate images (marker with 1050 nm LED and QD with 470 nm LED), which are later merged during image processing (see Supporting Information for images and histograms of uncertainties). With this approach, the contrast of the QDs and markers can be optimized independently to reduce the uncertainty in the final QD position. Consequently, the QD location uncertainty reduces significantly. However, the marker contrast is still limited by the background reflection of the illumination LED. In addition, one issue with this approach is that it is susceptible to image drift errors caused by the transition between two LEDs of different colors, which could potentially lead to larger localization errors. Alternatively, a second set of images (dark marker) is acquired with only the 470 nm LED to capture the bright QDs with a dark marker image. This approach involves leveraging the PL emission of the wetting layer to generate a dark image of the Au markers combined with the QD PL emission (see Supporting Information for images and histograms of uncertainties). This is achieved using a 780 nm LED for QD excitation to maximize the wetting layer emission. The enhanced wetting layer emission increases the contrast of the alignment markers, resulting in a dark image representation of the markers. Consequently, this approach achieves good localization accuracy for the markers. However, the decrease in the signal-to-noise ratio caused by the wetting layer emission increases the uncertainty in the QD location. Our analysis reveals that the Single-image 1 configuration has the best balance in terms of the accuracy of both QD and marker locations. However, it is noteworthy that the effectiveness of different image-acquisition approaches may vary depending on the design structure of the QD sample. The reflectivity contrast between the sample and the marker under the illumination wavelengths plays a significant role in determining the marker imaging quality, which subsequently influences marker location accuracy. Therefore, conducting reflectivity measurements for both the sample and the marker across a broad spectral range is advisable. This can help in selecting the optimal wavelength for the illumination LED to maximize the imaging contrast and, consequently, achieve the best accuracy in both QD and marker localization.\\

\noindent \textbf{QD localization via CL imaging}

Similarly, we use the CL imaging technique to determine the coordinates of the same set of QDs as in the PL technique. The setup used for the CL imaging of the combined marker and QDs fields is based on a Raith eLine Plus EBL system (see Methods section and Supporting Information). This state-of-the-art system can, in addition to performing high-resolution EBL, simultaneously detect secondary electrons and spectra pixel by pixel. Therefore, it enables us to obtain a clear SEM picture whilst also detecting the corresponding CL emission spectra of the sample. This results in a perfect overlap between the SEM picture and the obtained CL map (Figure \ref{figure 2} (e)). We would like to note that the CL map shows an intensity maximum in the top center, which is not caused by the lateral distribution QD emission intensity. This maximum is due to an efficiency optimum of our CL-mirror adjustment and is visible in all CL maps. The location of the QDs in the CL map is then determined via a 2D Gaussian fit (Figure \ref{figure 2} (f)) analog to the PL method. Based on a nonlinear least-squares approach, the peak position and one standard deviation as the fitting error is extracted. This way, it is possible to select the wavelength range for each QD individually and optimize the 2D fit accuracy. Afterward, the QD coordinates are uploaded into a python program, which extracts the center of the markers and transforms the QD coordinates into local coordinates, again with the top left marker as the origin. The center of the markers is hereby determined by a line scan across the arms of the markers followed by a straight line fit through the center of the arm. The resulting intercept of the two lines is regarded as the center. The fit is performed several times, and the error is the standard deviation based on this fitting series. Due to the positioning of the detector for the secondary electrons at the side of our system, the detection of these electrons is not uniform, leading to a brighter edge of the marker arm on one side and a darker edge or shadow on the other. The result is an asymmetry in the line scan profile, easily visible in Figure \ref{figure 2} (g), and based on this, a higher fit uncertainty in X-direction ((Figure \ref{figure 2} (h)). The measurements are performed at low temperatures of $20$ K under an acceleration voltage of $20$ kV. In each CL map, an area of (80 x 80) $\mu$m$^2$ is scanned by the electron beam with a pixel size of $500$ nm, and an exposure time per pixel of only $20$ ms, which highlights the high light throughput and a corresponding spatial SE image resolution of $25$ nm. Under those conditions, we achieve a mean uncertainty for the QD position of (8.09 $\pm$ 3.90) nm and (2.05 $\pm$ 0.54) nm for the alignment marker position, leading to a combined uncertainty (QD+marker) of (8.45 $\pm$ 3.75) nm (Table \ref{Table 1}).

\renewcommand{\arraystretch}{1.3}
\begin{table}
\caption{\text{Localization uncertainties for different positioning approaches.}}
	\centering 
	\begin{tabular}{l c c c c c}
        \hline \hline
        
		   & \multicolumn{1}{c}{\begin{tabular}[c]{@{}c@{}}QD location \\ uncertainty (nm)\end{tabular}}  & \multicolumn{1}{c}{\begin{tabular}[c]{@{}c@{}}Marker location\\ uncertainty (nm)\end{tabular}} & \multicolumn{1}{c}{\begin{tabular}[c]{@{}c@{}}QD + marker \\ uncertainty (nm)\end{tabular}} \\ 
           \hline
          
		 PL imaging: Single-image 1    & \begin{tabular}[c]{@{}l@{}} 1.38 $\pm$ 1.07 \\\end{tabular}  & 4.62 $\pm$ 1.65 & \begin{tabular}[c]{@{}l@{}} 4.51 $\pm$ 1.25 \\  \end{tabular}  \\
         
                               \hline
		 PL imaging: Single-image 2    & \begin{tabular}[c]{@{}l@{}} 2.57 $\pm$ 0.76 \\ \end{tabular}   & 4.16 $\pm$ 0.21 & \begin{tabular}[c]{@{}l@{}} 4.92 $\pm$ 0.42 \\  \end{tabular} \\
              \hline
                                                    
		 PL imaging: Merged-images    & \begin{tabular}[c]{@{}l@{}}0.93 $\pm$ 0.54 \\  \end{tabular}   & 5.12 $\pm$ 1.45 & \begin{tabular}[c]{@{}l@{}}5.80 $\pm$ 1.86 \\  \end{tabular} \\
              \hline
                
		 PL imaging: Dark-marker images   & \begin{tabular}[c]{@{}l@{}} 2.76 $\pm$ 1.57 \\  \end{tabular}   & 5.75 $\pm$ 0.89 & \begin{tabular}[c]{@{}l@{}} 6.41 $\pm$ 1.08 \\  \end{tabular}  \\
              \hline
		      
        CL imaging    & \begin{tabular}[c]{@{}l@{}} 8.09 $\pm$ 3.90 \\ \end{tabular}  & 2.05 $\pm$ 0.54 & \begin{tabular}[c]{@{}l@{}} 8.45 $\pm$ 3.75 \end{tabular}   \\
        \hline
	
        In-situ e-beam lithography    & \begin{tabular}[c]{@{}l@{}}  8.68 $\pm$ 2.86 \end{tabular}  & - & \begin{tabular}[c]{@{}l@{}} -  \end{tabular}   \\
        \hline
	
        In-situ e-beam lithography (Ref.\cite{Gschrey2015ResolutionFabrication})    & \begin{tabular}[c]{@{}l@{}} $25$ \end{tabular}  & - & \begin{tabular}[c]{@{}l@{}} -  \end{tabular}   \\
	
         \hline
         In-situ photolithography (Ref.\cite{Dousse2008})     & \begin{tabular}[c]{@{}l@{}} $50$ \end{tabular}  & - & \begin{tabular}[c]{@{}l@{}} -  \end{tabular}   \\
	
        \hline\hline
        \end{tabular}
\label{Table 1}
\end{table}

\section{QD-Mesa final alignment analysis}
\begin{figure}
\centering\includegraphics[width=1\textwidth]{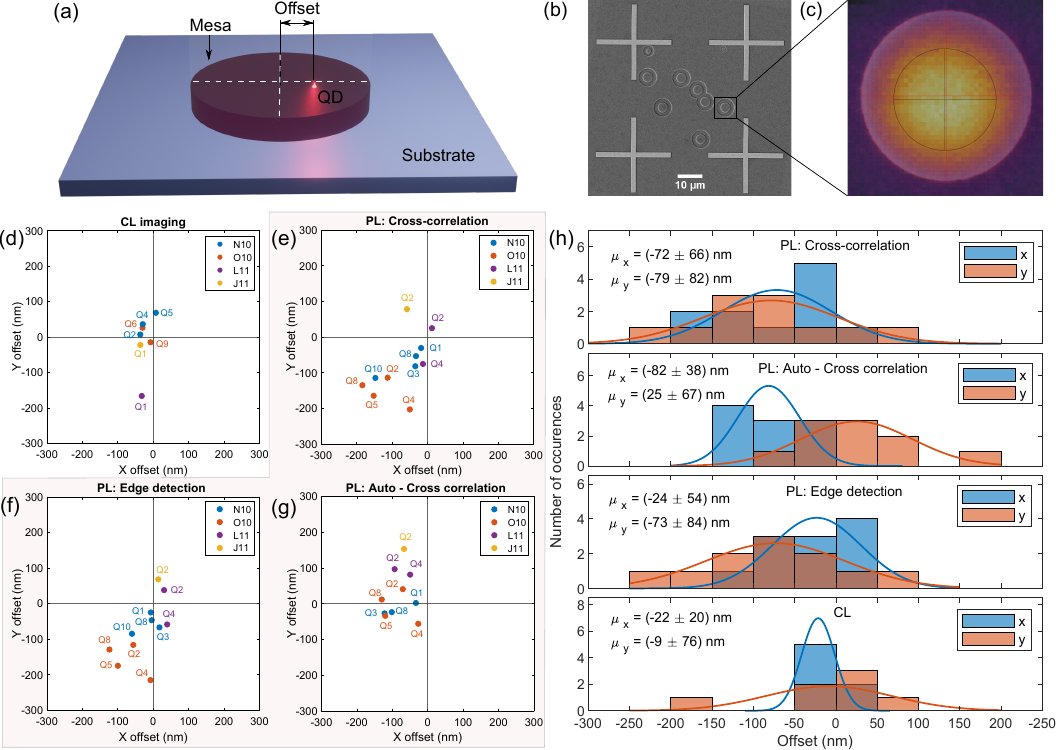}
\caption{Finding the accuracy of the determined QD location after nanostructure fabrication. (a) 3D sketch illustration of the QD misalignment relative to the center of the mesa structure. (b) SEM image of fabricated mesas around determined QDs coordinates. (c) CL map and SEM of QD in mesa structure, showing the QD emission profile within the mesa structure (QD center obtained from 2D Gaussian fit, and the mesa edges fitted with an ellipse).  QD offset distribution around the mesa center of CL imaging method in (d), and PL imaging (Single-image 1) method (shaded region) using cross-correlation, edge detection marker, and auto-cross correlation localization approaches in (e), (f), and (g), respectively. (h) Histogram of the offsets' distribution of all the methods with their mean offsets and standard deviation (uncertainty). N10, O10, L11, and J11 are the fields where the QDs have been located.}
\label{figure 3}
\end{figure}

Given that the fit uncertainties merely provide an estimate of the QD localization process accuracy and do not precisely reflect the true accuracy of QD location, we proceed to fabricate mesa structures (1.4 - 4.0 $\mu m$ in diameter) around the determined QD locations. This allows us to investigate the QD's true position uncertainty and its variance with the different positioning techniques. The position of the QD within the mesa is obtained by recording the SEM picture of the mesa and the QD CL map simultaneously. Due to the rather large dimensions of the mesa structures, the QDs are distant enough from the edges to prevent unwanted interactions between emitters and the edged surface. This way, we can precisely visualize the QD position within the mesa structure to retrieve the true location accuracy. In the resulting overlapped image (Figure 3 (b) and (c)), the edges of the mesa are fitted with an ellipse to identify the center of the mesa, and a Gaussian fit to the QD emission allows us to determine its position relative to the center of the mesa structure. Both fits deliver separate fitting errors, which, through Gaussian propagation, lead to an error for the final position offset. The mean value of this error for all mesa structures is (2.94 $\pm$ 0.90) nm and, therefore, is the precision of our fitting method. The small error shows the overall consistency of our fitting method. The offset of QDs from the center of the mesa for the CL and PL imaging techniques is shown in Figure \ref{figure 3} (d) \& (e). The result shows an offset tendency for both methods toward the -X axis; however, the PL method has a higher offset along the -X and an offset tendency toward the -Y axis. The CL method shows data points more evenly distributed around $\pm$Y. However, it should be noted that there are fewer data points for the CL compared to the PL method. Therefore, it is not possible to conclude with certainty that the CL has no tendency to a particular direction in Y. The tendency of the CL method to have a shift in the -X direction could be due to the previously discussed unequal illumination of the marker and resulting asymmetry in the marker scan in the X direction. This could lead to a line fit that is not well located in the center of the marker arm and, therefore, introduces a shift towards the negative X direction. Moreover, the source of these offsets includes the error from the EBL fabrication and the individual localization errors. Therefore, the source of these large offset is determined by obtaining the fabrication alignment error and analyzing the errors in the localization processes.

To investigate why the PL method has larger offsets and the reason it tends more to -X/-Y direction, we reanalyze our images with the different image processing algorithms we employed and compare the final alignment offsets (see Supporting Information for details of the different methods). First, We observe that for a QD that is not far off from a Gaussian profile, there is a small offset difference between the initial Gaussian blob detection and the subsequent MLE process (see Supporting Information). This suggests the presence of a small error source and indicates a lower contribution of the QD localization error to the overall alignment offset. This implies that the large offset values could be due to errors in the marker detection process. We investigate that by analyzing the QD offsets using different marker localization approaches, as shown in Figure \ref{figure 3}(f)-(g). The large offset difference between the various marker localization approaches indicates the presence of a bigger error source in the process. To identify the underlying problem, we first analyze the QDs with the largest offset for the cross-correlation method and observe that the large offsets in the negative X direction (mainly from field O10) result from a QD located along the marker edges (see Supporting Information Fig. S8). The cross-correlation algorithm employed is highly sensitive to the quality of the marker itself. This is because the overlap between the ideal and the image marker depends on the individual pixels within the cropped image area, meaning that any bright spot (QD) or artifact along the marker image area will result in an error that translates to an offset in the final alignment position. An alternative approach using the edge detection technique utilizes a line detection algorithm to locate the center of the marker. This method reduces reliance on individual pixels and minimizes the impact of bright spots or artifacts, providing a less sensitive and more robust marker localization approach. This is evident from the lower final alignment offsets, as seen in Figure \ref{figure 3}(f). Nonetheless, sharp marker edges are also critical for accurate localization using this approach. Another approach based on the cross-correlation method reported in \cite{Liu2017} involves taking the difference between the auto-correlation of the ideal marker with the cross-correlation of the ideal marker and marker image. This approach shows lower uncertainty in the QD position; however, it has a bigger offset in the X-axis, as shown in Figure \ref{figure 3}(g). The offsets distribution of all the techniques are summarized in \ref{figure 3}(h). Each of these approaches shows a different offset tendency. The reason for this difference is not very obvious and requires a deeper investigation into each approach. The CL imaging technique shows slightly lower offset and uncertainty values than the PL imaging techniques. This could be partly due to the more resolved marker image in the CL map, which results in more accurate marker localization. 

While we can see different offset values for the different techniques, it is unclear how much is either from the localization error or the fabrication alignment. To evaluate this, we characterize seven etched holes that were patterned simultaneously with the mesas. These holes are strategically positioned at the centers of optical fields in various corners of the sample area. The EBL error is determined by measuring the deviation between the actual position of each hole and the center of the corresponding markers' field using the same image processing approach (see Supporting Information Fig. S10). We find that each hole has a different deviation due to the rotational misalignment during the alignment procedure. A mean offset of $\approx$ (17 $\pm$ 22) nm and (-35 $\pm$ 14) nm from the corners within the located QDs' fields are obtained in the X and Y axes, respectively. That is, on average, the EBL alignment process tends to move the mesa position away from the QD by 17 nm in -X and 35 nm in +Y (QD center as the origin). It is noteworthy to point out that these values are estimates of the actual EBL offset to help us investigate the contribution of the EBL misalignment to the final position accuracy. A more detailed analysis of the EBL alignment rotation is needed for an accurate description. Figure \ref{figure 4} (a) shows the offset and uncertainty of both approaches after the EBL misalignment is compensated. The result implies that for both techniques, the offsets in -X result from errors in the localization process. For the PL imaging, almost half of the offset in -Y results from EBL misalignment. However, for the CL imaging technique, the offset tendency towards -Y results from the large EBL offset in -Y, and in fact, the localization process results in an offset tendency towards +Y values. Furthermore, the uncertainty in the offset values of all the techniques suggests a lower contribution of the EBL error in the final position uncertainty. \\ 

\begin{table}
\caption{Comparison of QDs positioning mean offsets and uncertainties for different positioning techniques}
	\centering 
	\begin{tabular}{ l c c c }
        \hline\hline

		   & \multicolumn{1}{c}{\begin{tabular}[c]{@{}c@{}} Positioning \\ offset in X \end{tabular}}  & \multicolumn{1}{c}{\begin{tabular}[c]{@{}c@{}}Positioning \\ offset in Y \end{tabular}} & \multicolumn{1}{c}{\begin{tabular}[c]{@{}c@{}} Reference \end{tabular}} \\
            \hline

		 PL imaging     & \begin{tabular}[c]{@{}l@{}}\hspace{0.05cm} $(-9 \pm {46})$ nm \\ $(-24 \pm {54})$ nm \end{tabular}   & \begin{tabular}[c]{@{}l@{}} \hspace{0.65cm} $-$ \\ $(-73 \pm {84})$ nm \end{tabular} & \begin{tabular}[c]{@{}l@{}} \hspace{0.45cm} Ref.\cite{Pregnolato2020DeterministicDots} \\(This work)\\  \end{tabular} \\
        \hline

		 CL imaging     & \begin{tabular}[c]{@{}l@{}} $(-22 \pm {20})$ nm \end{tabular}  & \hspace{0.05cm} $(-9 \pm {76})$ nm & \begin{tabular}[c]{@{}l@{}} (This work) \\  \end{tabular}  \\
           \hline

		 In-situ EBL     & \begin{tabular}[c]{@{}l@{}} \hspace{0.15cm} $(20 \pm {40})$ nm \end{tabular}   & \begin{tabular}[c]{@{}l@{}} $(-14 \pm {39})$ nm  \end{tabular} & \begin{tabular}[c]{@{}l@{}} (This work) \\  \end{tabular} \\

        \hline\hline
        \end{tabular}
	
\label{Table 2}
\end{table}

\section{QD localization and sample structuring via in-situ electron beam lithography:}

Furthermore, to compare the marker-based approach with the marker-free approach, we use the in-situ EBL technique to simultaneously locate the QDs and structure mesas around the QDs. In this case, the selected QDs are on a different sample field than those used for the marker-based techniques. By integrating CL spectroscopy and EBL within a single setup (detailed in the methods section), in-situ EBL offers a much simpler and time-efficient alternative to the imaging-based procedures since no additional alignment markers are required, and potential uncertainties due to the marker fitting process are eliminated. Moreover, possible imaging errors in the optical system are also eliminated. However, the process is more susceptible to drifts of the cryostat's cold finger (and thus the sample)  at low temperatures. An additional challenge lies in the requirement for the sample to be spin-coated with an EBL-resist during QD preselection via CL mapping. This introduces restrictions for the exposure time and the overall handling of the sample, an issue which, however, can be tackled by using machine learning enhanced in-situ EBL \cite{Donges2022MachineNanostructures}. The so-achieved uncertainty for the QD position is $8.68$ nm based on the same 2D Gaussian fit process described for the CL imaging. \\

\begin{figure}
\centering\includegraphics[width=0.7\textwidth]{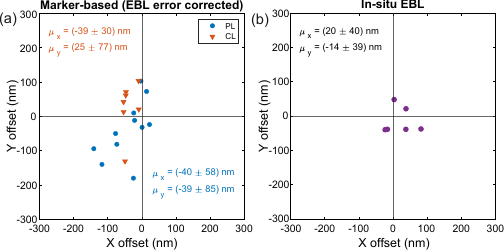}
\caption{Comparing the accuracy of marker-based positioning technique with in-situ EBL technique. (a) Comparison plot of the QDs offset distribution around the mesa center for marker-based PL (edge detection) and CL imaging after EBL error is compensated (to remove user-dependent error). (b) QDs offset distribution around the mesa center for the in-situ EBL technique.}
\label{figure 4}
\end{figure} 

The alignment accuracy of the in-situ EBL technique is analyzed identically to the marker-based methods, and the results are shown in Figure \ref{figure 4} (b). Notably, the in-situ EBL technique demonstrated smaller offsets and uncertainty values in comparison to the EBL-compensated marker-based methods. This can be attributed to its marker-free nature, effectively mitigating the substantial errors originating from the localization process. Nonetheless, an inclination towards an offset in the +X and -Y directions is observed. The source of this offset is attributed to the drift of the cryostat's cold finger at low temperatures. The whole process of CL mapping, QD fitting, and EBL takes up to 3 minutes, which is time enough to lead to a $20 - 30$ nm drift (as we determined independently) and, therefore, can explain a majority of the offset. For future considerations, a strategic approach to achieve better alignment accuracy could involve actively monitoring and compensating for this cryostat drift during the in-situ EBL nanofabrication procedure.

\section{Discussion}

In the realm of QD positioning research, as far as our knowledge extends and despite the fact the knowledge about the alignment accuracy is a key parameter of deterministic nanoprocessing, prior to our work, only Pregnolato et al. reported the final alignment offsets and uncertainties after QD integration into nanostructures using the PL imaging approach \cite{Pregnolato2020DeterministicDots}. However, their structures were smaller than the QD emission profile, and the optical diffraction limit constrained the positioning accuracy. In our systematic study, we implemented a more robust approach by fabricating mesa structures of sufficient size to contain the QD emission profile and utilized a high-resolution CL mapping technique to extract the accurate positions of the QD within the mesas. We investigated and compared the final alignment accuracy of the QDs to the mesa structures using marker-based PL and CL imaging techniques. Both methods yielded an average localization uncertainty of $<10$ nm, consistent with previous studies. However, we encountered significant alignment offsets and uncertainties following the alignment of mesa nanostructures to the localized QDs. Notably, both approaches exhibited tendencies for offsets along specific axis directions, with a final alignment uncertainty of $<100$ nm for both techniques. \rev{This result is in good agreement with previous findings \cite{Gschrey2013}, including results based on polarization measurements \cite{Peniakov2023PolarizedResonator}, as well as comparisons between Purcell factor simulations and experiments results \cite{Sapienza2015,Kolatschek2019DeterministicLithography}. This implies that these uncertainties are not specific to our setup but are current limitations of the approaches studied.} 

The considerable discrepancy between the localization uncertainty and the final alignment uncertainty, which is the crucial figure for the later device performance, suggests the presence of larger unaccounted errors arising from the localization and fabrication processes. This finding raises concerns about the reliability of the localization fit uncertainties obtained in this work and prior studies. We found that the main contributor to the large mean offset in the PL imaging technique was the marker localization process. Thus, the low uncertainty obtained from the fitting functions does not equate to the QD location accuracy, as the uncertainty is also limited by the imaging system's resolution (pixel size). Table \ref{Table 2} summarizes the offset and uncertainties of the different QD positioning approaches. \rev{The CL imaging approach shows better accuracy due to the better marker image contrast compared to the PL image, which has blurry marker edges because of a worse reflectivity contrast between the marker and the background.} While the marker-based approaches are simpler to implement, the errors that arise from the marker localization result in lower position accuracy with offset tendencies to a certain direction. \rev{Therefore, a highly resolved image of the marker is needed for more accurate localization. For the CL approach, this could be achieved by reducing the overall size of the marker area. This way, the pixel size of 500 nm used could be reduced without increasing the scanning time and thus obtain a higher-resolution image. For the PL approach, better accuracy could be achieved by employing better illumination LEDs at different wavelengths to optimize the reflectivity contrast between the marker and the background, leading to sharper marker edges. Furthermore, alternative marker geometries could be explored. A design choice with a more narrow diameter at the very center of the marker could lead to an improved detection of the center. This could be complemented by a new image-processing approach, which could be employed to detect the position of the markers with high accuracy, even with low-quality markers.} In addition, a more precise fabrication alignment is necessary for accurate positioning. Though the accuracy varies from user to user, a more automated approach would be more reliable for high reproducibility. \rev{A possible approach could be supported using machine learning-enhanced image processing.} For the in-situ EBL process, we see an improvement in the alignment accuracy, which we attribute to the simpler process flow without needing alignment markers. We assume that the offset here is caused mainly by temperature-induced drifts of the cryostat's cold finger, and therefore, compensating for those drifts could be an easy approach to further improve the alignment accuracy in the future. \rev{This could be achieved by taking a defect or prefabricated marker on the sample surface as a reference point and using it after every scanned line to readjust the starting point for the mapping procedure.} Although we obtained comparable results for all approaches, they each have unique advantages and challenges. While the marker-based approaches require marker fabrication steps and extensive pre-characterization, the in-situ EBL technique enables simple pre-selection of QDs during a prior CL mapping step in an easy-one-coordinate system. In return, the dwell time per pixel during the mapping procedure is strongly constrained since the sample is already coated with a resist, a problem that can be mitigated by using in-situ EBL with machine learning \cite{Donges2022MachineNanostructures}. Still, the flexibility in terms of illumination time during optical imaging provides an essential advantage for the marker-based methods, especially in the case of darker QDs (e.g. in the telecom O- and C-band), and additionally, the PL-based approach is more flexible in the choice of different excitation schemes.  

\begin{table}
\caption{\text{\rev{Performance of the different photonic structures as a function of QD displacement from center}}}
	\centering 
	\begin{tabular}{ l c c c }
        \hline\hline
        
		   & \multicolumn{1}{c}{\begin{tabular}[c]{@{}c@{}} QD displacement \\in X \&Y(nm) \end{tabular}}  & \multicolumn{1}{c}{\begin{tabular}[c]{@{}c@{}} $\%$ of maximum \\PEE/CE\\ in X(Y) axis \end{tabular}} & \multicolumn{1}{c}{\begin{tabular}[c]{@{}c@{}} $\%$ of maximum \\ Purcell factor\\ in X(Y) axis \end{tabular}} \\
            \hline

		 Micropillar (2.1 $\mu$m) \textsuperscript{This work}     & \begin{tabular}[c]{@{}l@{}} $100$  \\ $200$  \end{tabular}   & \begin{tabular}[c]{@{}l@{}}  $99$($99$)$\%$ \\ $99$($99$)$\%$ \end{tabular} & \begin{tabular}[c]{@{}l@{}}  $98$($98$)$\%$ \\ $92$($92$)$\%$ \end{tabular} \\
        \hline
		 Bullseye \textsuperscript{This work}     & \begin{tabular}[c]{@{}l@{}} $50$  \\ $100$ \\$200$ \end{tabular}   & \begin{tabular}[c]{@{}l@{}}  $82$($47$)$\%$ \\ $44$($4$)$\%$ \\$2$($56$)$\%$ \end{tabular} & \begin{tabular}[c]{@{}l@{}}  $96$($95$)$\%$ \\$84$($19$)$\%$ \\ $8$($96$)$\%$ \end{tabular} \\
        \hline
		Nanobeam cavity \cite{Li2023ScalableCircuit}     & \begin{tabular}[c]{@{}l@{}} $50$  \end{tabular}   & \begin{tabular}[c]{@{}l@{}}  $95(70)\%$   \end{tabular} & \begin{tabular}[c]{@{}l@{}}  $85(25)\%$   \end{tabular} \\
        \hline

		Photonic crystal waveguide \cite{Uppu2020On-chipSource}     & \begin{tabular}[c]{@{}l@{}} $50$ \\ $75$  \end{tabular}   & \begin{tabular}[c]{@{}l@{}}  $75(-)\%$ \\ $50(-)\%$   \end{tabular} & \begin{tabular}[c]{@{}l@{}}  $-$ \\ $-$  \end{tabular} \\

        \hline\hline
        \end{tabular}
\label{Table 3}
\end{table}

\rev{Nevertheless, the best measure of the reliability of these techniques for high-yield fabrication of high-performance single-QD-based quantum devices and for future on-chip quantum photonic circuits lies in ensuring that offset values fall within a range at which the performance of the individual units is maintained to an acceptable value. To discuss to what extent our methods can be used for the fabrication of high-performance single-photon source (SPS) devices, we performed simulations 
on the efficiency and Purcell factor as a function of the QD displacement in the two most prominent SPSs, e.g., micropillar and CBG. The micropillar design has shown little sensitivity with a negligible drop of $<2\%$ in both the Purcell factor and photon extraction efficiency (PEE) within 100 nm of QD displacement in both X and Y directions (see supporting information Figure S11). The CBG (bullseye) design has proven to be more sensitive due to the small central disc diameter (see supporting information Figure S12). For a QD displacement of 50 nm, a significant purcell factor drop of $\approx 50\%$ is observed in the Y direction, compared to $\approx 20\%$ in the X direction. This large difference highlights the significant challenge of achieving high polarization control in these devices. However, the PEE remains within a $5\%$ change for a 50 nm displacement. Moreover, on-chip integration of QDs in nanobeam cavities/waveguides using the in-situ EBL has recently been studied. The reported values show a coupling efficiency (CE) of $>29\%$, and a Purcell factor of $\sim 1.8$ is preserved for up to a 50 nm QD offset\cite{Li2023ScalableCircuit}. Table \ref{Table 3} summarizes the performance of the various photonic structures for different ranges of QD displacement.} \rev{Thus, the alignment accuracy of current techniques is not yet sufficient for the scalable fabrication of photonic circuits. This is a significant challenge for scalable quantum information technologies and should be the focus of future efforts in the community.}

\section{Conclusion}

In conclusion, we have studied and compared the overall alignment offset and uncertainty of a QD integrated into a circular mesa structure using PL imaging, CL imaging, and in-situ EBL positioning techniques. Our results revealed that the localization accuracy of the marker-based techniques, given by the fit errors of the QD position, does not fully represent the accuracy of QD integration process. This conclusion is drawn from the observed mean offsets and large uncertainty in the final QD position within the mesa structures, which are significantly larger than the accuracy of the QD position obtained in the preselection. These inaccuracies primarily stem from the presence of unaccounted errors during the marker localization process and the EBL fabrication alignment. On the other hand, the in-situ EBL technique demonstrated comparable accuracy of the QD position but better final QD alignment accuracy with the mesa structure, primarily because it does not rely on markers for positioning. Our study is a crucial step in understanding the optimal approach for high-throughput fabrication of highly efficient single-QD-based quantum devices, which is essential for the advancement of scalable photonic quantum information technologies. By shedding light on the limitations and strengths of different positioning techniques, our research contributes to the further development of advanced QD integration techniques for QD-based photonic structures.

\section{Methods}
\label{section:methods}

\noindent \textbf{Sample preparation:}

The QD sample is grown on an undoped GaAs wafer using molecular beam epitaxy (MBE). It consists of low-density InAs QDs embedded in a GaAs membrane with a thickness of 242.4 nm, followed by a 1.5 $\mu$m thick Al0.9GaAs grading layer and a 300 nm buffer layer. The alignment markers are fabricated through standard EBL and lift-off processing. The sample is spin-coated with a positive electron-beam resist (CSAR AR-P 6200.09), which is exposed with an EBL machine (JEOL 9500) and developed with n-Amyl acetate (ZED). The sample is then deposited with 5/50 nm titanium/gold using an electron-beam evaporator. The Ti/Au and resist in the unexposed area are lift-off using Microposit Remover 1165 with a gentle sample agitation and then rinsed with Acetone, IPA, and DI water.\\

\noindent \textbf{QD localization: PL} 

After the fabrication of the alignment markers, the sample is mounted in a closed-cycle cryostat operating at 4 K (Attodry). The cryostat is equipped with a piezoelectric positioning stage and a low-temperature microscope objective (magnification = 60$\times$, NA = 0.82) located inside the cryostat. Low-temperature PL imaging of the QDs is performed to determine the spectral and spatial position of the individual QDs. The micro-PL setup is based on the two-color PL imaging technique, which utilizes two LEDs at 470 nm and 1050 nm to image the QD PL and marker, respectively (see Supporting Information for setup sketch). All the sets of images are acquired and analyzed with an image analysis program (developed with Python, details in Supporting Information), and the individual QDs' positions with respect to the alignment markers are extracted. Different image acquisition approaches are employed to investigate the most accurate technique. Cross-correlation and edge detection algorithms are employed for the marker localization, while Gaussian blob detection combined with a maximum likelihood estimation algorithm is employed for the QD localization. The uncertainties in locating the actual position of the QDs for the subsequent mesa fabrication are obtained from the fit uncertainties of the QD and marker intensity profiles.\\

\noindent \textbf{QD localization: CL} 

The setup used to perform the CL measurements is based on a Raith eLine Plus electron beam lithography system equipped with a CL extension. It includes a He-flow cryostat to enable low-temperature measurements, and the CL emission is collected by a parabolic mirror (NA = 0.88) and is directed into a spectrometer. A Si-CCD and a 1D InGaAs diode array are used to cover the spectral range of $300$ - $1700$ nm. Whilst scanning a marker area, we obtain simultaneously CL and SEM data through a secondary electron detector. As the next step we determined via a 2D Gaussian fit the position of the QDs in the CL map (LabView) and uploaded the coordinates into an image analysis program (Python) which fits the center of the markers through line scans and transforms the QD coordinates into local coordinates. All uncertainties are obtained through the fit uncertainties of the CL and SEM images.\\

\noindent \textbf{Mesa patterning:}

After the central position of the QDs is determined, the coordinates are transformed into the global sample coordinates, and mesa structures are fabricated around the transformed preselected coordinates. The mesa is fabricated through EBL and an etching process. The sample is spin-coated with a positive tone electron-beam resist (CSAR AR-P). The exposure is done with a 100 keV EBL machine (JEOL 9500) with a 6 nA current. The mesa structures are aligned to the sample using the JEOL EBL automatic alignment procedure with two diagonal P and Q markers. The resist was developed with AR-600 developer and the pattern was defined through ICP-RIE etching down to a depth of approximately 230 nm.\\

\noindent \textbf{EBL alignment error:}

The EBL error was measured by characterizing 7 etched holes patterned in the same lithography step as the mesas. The holes were positioned at the centers of the optical fields in various corners of the sample. To determine the spatial variation in the EBL error (taken as the deviation in the hole's actual position from the center of the field), an SEM image of the different fields is taken. The center locations of the markers and holes are found using the same image analysis program. The misalignment offset of the holes from the center of the four markers field is obtained as the EBL error (See Supporting Information for details).
\\

\noindent \textbf{QD positioning: in-situ EBL} 

For the QD positioning through in-situ EBL we used the same setup as for the CL marker method. The sample is first spin coated in a clean room facility with the resist CSAR 6200.13 at 6000 rpm. It is followed by the main process consisting of three consecutive steps. First, a CL map scan of a plain area on the sample is performed with a map size of (20 x 20) $\mu$m$^2$, a pixel size of 500 nm and an exposure time of 30 ms. Second, the position of a desired QD is determined through a 2D Gaussian fit (LabView). Third, the preselected QD is integrated via EBL into a mesa structure by utilizing the negative-tone regime of the EBL resist.

\section*{Associated Content}
\noindent
\textbf{Supporting Information}

\noindent
The Supporting Information is available free of charge at \url{https://pubs.acs.org/doi/10.1021/acsphotonics.xxxxx.}

\rev{PL and In-situ/CL imaging setups sketch, markers localization image analysis procedures, position uncertainties from other localization techniques, camera images and position uncertainties of other PL imaging approaches (Single-image 2, Merged images, and Dark marker image), final position uncertainties of other PL imaging approaches, processed markers crop images of the different optical fields, QD offsets distribution from QDs localized using Gaussian blob detection, EBL offset at different sample regions, efficiency and Purcell factor simulations for different QD displacement.}\\

The Python script used for the PL image analysis is available at \url{https://doi.org/10.5281/zenodo.10376526}. 

\section*{Funding}

The authors acknowledge the European Research Council (ERC-CoG "Unity", grant no.865230) and support from the Independent Research Fund Denmark (Grant DFF-9041-00046B). N.G. acknowledges support from the European Union's Horizon 2020 Research and Innovation Program under the Marie Skolodowska-Curie Grant Agreement no. 861097. B.M. also acknowledges support from the European Research Council (ERC-StG "TuneTMD", grant no. 101076437), and the Villum Foundation (grant no. VIL53033). S.L., H.L., D.D., X.Su., X.Sh., H.N., and Z.N. acknowledge support from National Key Technologies  R\&D Program of China, 2018YFA0306101, Key-Area Research and Development Program of Guangdong Province (Grant No. 2018B030329001) and National Natural Science Foundation of China, 62035017, 61505196.
J.D., S.L., J.S. and S.R. acknowledge funding support from the German Research Foundation (Nos. Re2974/25-1 and INST 131/795-1 320 FUGG), and via the SEQUME project (20FUN05) from the EMPIR program cofinanced by the Participating States and from the European Union’s Horizon 2020 research and innovation program. 

\section*{Notes}
The authors declare no competing financial interest.

\section*{Acknowledgement}
The authors thank Jonas Winther and Kasper Steinmuller for their helpful contributions to the image processing program.  

\section*{Author contributions}

A.M. and J.D. contributed equally to this work. S.L., H.L., D.D., X.Su., X.Sh., H.N., and Z.N. grew the QD samples. A.M. and B.M. performed photoluminescence imaging experiments. A.M., B.G., and B.M. contributed to image analysis and data processing. J.D., J.S., and S.Ro. performed the cathodoluminescence imaging and in-situ EBL experiments, including image analysis and data processing. \rev{M.A.J performed simulations.} N.G., S.R., and B.M. conceived the idea and coordinated the project. All authors wrote the paper.

\bibliography{citations}

\newpage

\section{Graphical TOC Entry}
 \begin{figure}
\includegraphics[width=0.5 \textwidth]{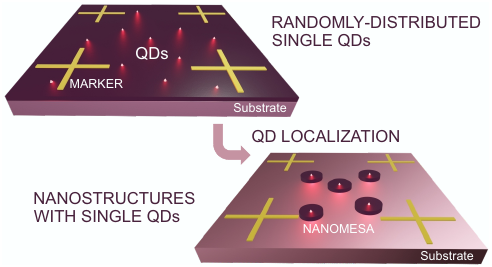}
\end{figure}

\end{document}